\documentstyle[aps,epsfig,preprint]{revtex}
\begin{document}
\draft

\title{\large Deformation effects in $^{56}$Ni nuclei produced in
$^{28}$Si+$^{28}$Si at 112 MeV } 

\author{C.~Bhattacharya{\thanks{Permanent address : VECC, 1/AF Bidhan Nagar,
Kolkata 64, India.}}, M.~Rousseau, C.~Beck{\thanks{Corresponding author.
Electronic address: christian.beck@ires.in2p3.fr.}}, V.~Rauch, R.~M.~Freeman,
D.~Mahboub{\thanks{Present address : University of Surrey, Guildford GU2 7XH,
United Kingdom.}}, R. Nouicer\thanks{Present address : University of Illinois
at Chicago, and Brookhaven National Laboratory, USA.}, P.~Papka, and
O.~Stezowski{\thanks{Permanent address : IPN Lyon, F-69622 Villeurbanne,
France.}}} 

\address{\it Institut de Recherches Subatomiques, UMR7500, Institut National de
Physique Nucl\'eaire et de Physique des Particules - Centre National de la
Recherche Scientifique/Universit\'e Louis Pasteur, 23 rue du Loess, B.P. 28,
F-67037 Strasbourg Cedex 2, France}

\author{A.~Hachem and E.~Martin}

\address{\it Universit\'e de Nice-Sophia Antipolis, F-06108 Nice, France}

\author{A.~K.~Dummer{\thanks{Present address : Triangle Universities Nuclear
Laboratory, University of North Carolina, Durham, NC 27708-0308, USA}} and
S.~J.~Sanders} 

\address{\it Department of Physics and Astronomy, University of Kansas,
Lawrence, Kansas 66045, USA}

\author{A.~Szanto De Toledo}

\address{\it Departamento de F\'{\i}sica Nuclear, Instituto de F\'{\i}sica da
Universidade de S\~ao Paulo, C.P. 66318-05315-970 - S\~ao Paulo, Brazil}

\date{\today}
\maketitle

\newpage

\begin{abstract}

{Velocity and energy spectra of the light charged particles (protons and
$\alpha$-particles) emitted in the $^{28}$Si(E$_{lab}$ = 112 MeV) + $^{28}$Si
reaction have been measured at the Strasbourg {\sc VIVITRON} Tandem facility.
The {\sc ICARE} charged particle multidetector array was used to obtain
exclusive spectra of the light particles in the angular range 15$^{\circ}$ -
150$^{\circ}$ and to determine the angular correlations of these particles with
respect to the emission angles of the evaporation residues. The experimental
data are analysed in the framework of the statistical model. The exclusive
energy spectra of $\alpha$-particles emitted from the $^{28}$Si + $^{28}$Si
compound system are generally well reproduced by Monte Carlo calculations using
spin-dependent level densities. This spin dependence approach suggests the
onset of large deformations at high spin. A re-analysis of previous
$\alpha$-particle data from the $^{30}$Si + $^{30}$Si compound system, using
the same spin-dependent parametrization, is also presented in the framework of
a general discussion of the occurrence of large deformation effects in the
A$_{\small CN}$ $\approx$ 60 mass region.} 

\end{abstract} 
{PACS number(s): 25.70.Gh, 25.70.Jj, 25.70.Mn, 24.60.Dr}

\newpage

\section{INTRODUCTION}

In recent years, there have been a number of experimental and theoretical
studies~\cite{sanders99} aimed at understanding the decay of light compound
nuclei (CN) and dinuclear systems (A$_{\small CN}$ $\leq$ 60) formed through
low energy heavy-ion reactions (E$_{lab}$ $\leq$ 10 MeV/nucleon). In most of
the reactions studied, the properties of the observed, fully energy damped
yields have been successfully explained in terms of either a fusion-fission
(FF) mechanism or a heavy-ion resonance
behavior~\cite{sanders99,zurmuhle83,betts81}. The strong resonance-like
structures observed in elastic and inelastic excitation functions of
$^{24}$Mg + $^{24}$Mg~\cite{zurmuhle83} and $^{28}$Si + $^{28}$Si
scatterings~\cite{betts81} have been suggestive of the presence of shell
stabilized, highly deformed configurations in the $^{48}$Cr and $^{56}$Ni N = Z
dinuclear systems, respectively~\cite{sanders99}. The investigation of the 
structure of the doubly-magic $^{56}$Ni nucleus is particularly interesting,
with the recent observation~\cite{rudolph99} in this system of deformed bands
that may be the precursors of large deformation or superdeformation behavior in
the A$_{\small CN}$ $\approx$ 60 mass region~\cite{svenson97,rudolph98}. 

In a recent experiment using the {\sc EUROGAM} phase II $\gamma$-ray
spectrometer, we have investigated~\cite{nouicer99,beck01} the possibility of
preferential population of highly deformed bands in the symmetric fission
channel of the $^{56}$Ni$^{*}$ CN, produced through the $^{28}$Si + $^{28}$Si 
reaction at E$_{lab}$ = 112 MeV, which corresponds to the energy of the
conjectured J$^{\pi}$ = 38$^{+}$ quasimolecular resonance~\cite{betts81}. 
Some evidence for this behavior was observed~\cite{nouicer99,beck01}, but
was not definitively conclusive~\cite{beck01}.

The present work involves the search for the possible occurence of highly
deformed configurations of the $^{56}$Ni$^{*}$ CN produced in the
$^{28}$Si + $^{28}$Si reaction. Light charged particles (LCP) emitted at the
resonance energy~\cite{nouicer99,beck01} of E$_{lab}$ = 112 MeV, and in-plane
coincidences of the LCP's with both evaporation residues (ER) and FF fragments
have been measured. The LCP's emitted during the CN decay processes carry
information on the underlying nuclear shapes and level densities. In
particular, new information on nuclear structure far above the yrast line can
be obtained from their study by a comparison with statistical model
calculations~\cite{viesti88}. The LCP's emitted from FF fragments may provide
the deformation properties of these fragments. Studies of nuclear shapes based
on evaporated LCP spectra have evoked considerable interest and controversy
~\cite{viesti88,govil87,larana88,huizenga89,agnihotri93,govil98,bandyopadhyay99,govil00a,govil00b,mahboub01}.
For example, an extremely large deformation suggested in the decay of the
$^{60}$Ni$^*$ CN~\cite{larana88}, formed in the reaction $^{30}$Si + $^{30}$Si
at E$_{lab}$ = 120 MeV, was not supported by the statistical model analysis of
Nicolis and Sarantites~\cite{nicolis89}. 

In this paper we will focus on the LPC's found in coincidence with ER's. These
data will be analysed with the {\sc CACARIZO} statistical model
code~\cite{viesti88}. Section II describes the experimental procedures. In Sec.
III we present the data analysis of the exclusive $^{28}$Si + $^{28}$Si data
(part of the experimental results presented here in detail have already been
briefly reported elsewhere
~\cite{bhattacharya99,rousseau00,beck00,bhattacharya00,rousseau01a}). The
statistical model calculations are compared to the experimental data in Sec.
IV. This section includes a re-analysis of existing $\alpha$-particle data from
the $^{30}$Si + $^{30}$Si reaction previoulsy measured by La Rana et
al.~\cite{larana88} using a consistent set of input parameters fitting both
$^{28}$Si + $^{28}$Si and $^{30}$Si + $^{30}$Si $\alpha$-particle spectra. We
end with a summary of our results in Sec. V. 

\newpage

\section{EXPERIMENTAL PROCEDURES}

The experiment was performed at the IReS Strasbourg {\sc VIVITRON} Tandem
facility, using a 112 MeV $^{28}$Si beam incident on a self-supported 180
$\mu$g/cm$^2$ thick $^{28}$Si target prepared at IReS. Additional targets of
natural $^{12}$C, gold, and Formvar were irradiated for calibration, background
determination, and normalization purposes. The elemental compositions of the
$^{28}$Si target was accurately determined at IReS using Rutherford back
scattering (RBS) techniques with $^{1}$H and $^{4}$He beams provided by the 4
MV Van de Graaff accelerator~\cite{beck00}. The main target contaminants were C
and O, each contributing less than 2$\%$ to the total number of atoms in the
target, and Cu. Because of the relatively low beam energy with respect to the
barrier energy for Cu, this contaminant is not expected to affect our results.
The natural C target was used to obtain the background correction for this
element. These corrections were found to be relatively small. 

Both the heavy ions and their associated LCP's were detected using the {\sc
ICARE} charged particle multidetector array~\cite{rousseau01b}. The heavy
fragments (ER, quasi-elastic, deep-inelastic and FF fragments) have been
detected in 6 gas-silicon hybrid telescopes (IC), each consisting of an
ionization chamber, with a thin Mylar entrance window, followed by a 500 $\mu$m
thick Si(SB) detector. The IC's were located at $\Theta_{lab}$ = -15$^\circ$,
-20$^\circ$, -25$^\circ$, -30$^\circ$, -35$^\circ$, and -40$^\circ$ in two
distinct reaction planes (for each plane, the positive and negative angles
refer to the opposite and same side of the beam as the heavy-ion IC detector,
respectively). The in-plane detection of coincident LCP's was done using four
triple telescopes (40 $\mu$m Si, 300 $\mu$m Si, and 2 cm CsI(Tl)) placed at
forward angles ($\Theta_{lab}$ = +15$^\circ$, +25 $^\circ$, +35$^\circ$, and
+45$^\circ$), 16 two elemental telescopes (40 $\mu$m Si, 2 cm CsI(Tl)) placed
at forward and backward angles (+40$^\circ$ $\leq$ $\Theta_{lab}$ $\leq$
+115$^\circ$), and finally two other IC's telescopes placed at the most
backward angles $\Theta_{lab}$ = +130$^\circ$ and +150$^\circ$. The IC's were
filled with isobutane and their pressures were kept at 30 Torr at backward
angles and 60 Torr at forward angles, respectively, for detecting heavy
fragments and light fragments. The acceptance of each telescope was defined by
thick aluminium collimators. 

The calibration of the {\sc ICARE} multidetector array was done using
radioactive $^{228}$Th $\alpha$-sources, a precision pulser, and elastic
scattering of 112 MeV $^{28}$Si from $^{197}$Au, $^{28}$Si, and $^{12}$C
targets in a standard manner. In addition, $\alpha$-particles emitted in the 
$^{12}$C($^{16}$O,$^{4}$He)$^{24}$Mg$^{*}$ reaction at E$_{lab}$($^{16}$O) = 53
MeV provide known $\alpha$-particle energies from the decay of $^{24}$Mg
excited states thus allowing for the calibration of the backward angles
detectors~\cite{beck00,rousseau01b}. The proton calibration was done using
scattered protons from Formvar targets using both the $^{28}$Si and $^{16}$O
beams. More details on the experimental setup of {\sc ICARE} and on the
analysis procedures can be found in Refs.~\cite{beck00,rousseau01a,rousseau01b}
and references therein. 

\newpage

\section {DATA ANALYSIS AND EXPERIMENTAL RESULTS}

The velocity contour maps of the LCP Galilean invariant differential
cross-sections (d$^{2}\sigma$/d$\Omega$dE)p$^{-1}$c$^{-1}$ as a function of the
LCP velocity provides an overall picture of the reaction pattern. Figs.~1.a)
and 1.b) show such two-dimensional scatter plots for $\alpha$-particles and
protons, respectively, measured in a single mode. For a sake of clarity the
velocity cuttoffs arising from the detector low-energy thresholds are indicated
for each telescopes. V$_{\parallel}$ and V$_{\perp}$ denote laboratory velocity
components parallel and perpendicular to the beam, respectively. Figs.~1.c) and
1.d) are the corresponding plots as calculated by the statistical model
discussed in a following section. The dashed circular arcs, centered on the
center of mass V$_{c.m.}$ and defined to visualize the maxima of particle
velocity spectra, describe the bulk of data rather well as can be observed for
instance for the proton energy spectra. They have radii very close to the
Coulomb velocities of $\alpha$-particles and protons in the decay of
$^{56}$Ni$^{*}$ $\rightarrow$ $^{52}$Fe + $^{4}$He, and of $^{56}$Ni$^{*}$
$\rightarrow$ $^{55}$Co + $^{1}$H, respectively. The agreement is apparently
deteriorated for the $\alpha$-particle spectra by the relatively large
low-energy thresholds of the most backward-angle telescopes. Despite these
deviations, the spectra can be understood by assuming a sequential evaporative
process and successive emission sources starting with the thermally
equilibrated $^{56}$Ni$^{*}$ CN until the final source characterised by a
complete freeze-out of the residual nucleus. The invariant cross section
contours fall around the dashed circular arcs centered at the CN recoil
velocity V$_{CN}$ = V$_{c.m.}$, and represent the isotropic emission patterns
to be expected for a fusion-evaporation mechanism after full-momentum transfer
and complete fusion (CF). LCP's emitted from direct reactions or from a
pre-thermalization emission process would have manifested themselves as even
stronger deviations from the dashed circular arcs, as shown at much higher
bombarding energies, E($^{28}$Si) = 12.4, 19.7, and 30 MeV/nucleon, for the
$^{28}$Si + $^{28}$Si reaction~\cite{decowski88,meijer91}. In these early
works~\cite{decowski88,meijer91} only a few pre-equilibrium LCP's have been
shown to be emitted prior to fusion at the lowest energy E($^{28}$Si) = 12.4
MeV/nucleon~\cite{decowski88,meijer91}, therefore the absence of a
pre-equilibrium component at the present energy E($^{28}$Si) of 4 MeV/nucleon
is expected. 

Figs.~2.a) and 2.b) show the exclusive scatter plots of Galilean invariant
differential cross-sections in the (V$_{\parallel}$,V$_{\perp}$) plane, for
$\alpha$-particles and protons measured in coincidence with all ER's (20 $\leq$
Z $\leq$ 25), identified in a IC detector located at $\Theta^{ER}_{lab}$ =
-15$^{\circ}$. Figs.~2.c) and 2.d) are statistical model predictions discussed
in Sec. IV. As for the inclusive data, there is no bias due to direct or
preequilibrium processes. Thus we are confident that essentially all the
emitted particle are associated with a statistical de-excitation process
arising from a thermalized source such as the $^{56}$Ni CN. The energy spectra
of these particles, presented thereafter, are also strongly supportive of this
conclusion. 

Typical exclusive energy spectra of the corresponding $\alpha$-particles are
shown in Fig.~3 at the indicated angles (from $\Theta^{LCP}_{lab}$ =
+40$^{\circ}$ to $\Theta^{LCP}_{lab}$ = +65$^{\circ}$) for the $^{28}$Si +
$^{28}$Si reaction at E$_{lab}$($^{28}$Si) = 112 MeV. The measured
$\alpha$-particles are in coincidence with all ER's (20 $\leq$ Z $\leq$ 25)
identified in the IC detector located at $\Theta^{ER}_{lab}$ = -15$^{\circ}$
(the IC's located at more backward angles have too low statistics for
fusion-evaporation events to be used in the analysis). The spectral shapes of
the coincident $\alpha$-particles are very similar to inclusive energy spectra 
but without low-energy, non-Maxwellian contributions, as shown in Fig.~6. All
the spectra have Maxwellian shapes with an exponential fall-off at high energy
which reflects a relatively low temperature (T$_{slope}$ $\approx$
[8E$^{*}_{\small CN}$/A$_{\small CN}$]$^{1/2}$ = 3.1 MeV) of the decaying
nucleus. The shape and high-energy slopes are also found to be essentially
independent of angle in the c.m. system. These behaviors suggest, as for the
velocity spectra, a statistical CN decay process.

The in-plane angular correlations of $\alpha$-particles and protons (measured
in the -115$^{\circ}$ $\leq$ $\Theta^{LCP}_{lab}$ $\leq$ +115$^{\circ}$ angular
range) in coincidence with all the ER's (20 $\leq$ Z $\leq$ 25), produced in
the $^{28}$Si(112 MeV) + $^{28}$Si reaction, are shown in Fig.~4. The angular
correlations are peaked strongly on the opposite side of the ER detector
located at $\Theta^{ER}_{lab}$ = -15$^{\circ}$ with respect to the beam. This
peaking is the result of the momentum conservation. The solid lines shown in
the figure are the results of statistical model predictions for CF and
equilibrium decay using the code {\sc CACARIZO}~\cite{viesti88}, as discussed
in the next section. 

\newpage

\section{STATISTICAL MODEL CALCULATIONS AND DISCUSSION}

The analysis of the data has been performed using {\sc
CACARIZO}~\cite{viesti88}, the Monte Carlo version of the statistical model
code {\sc CASCADE}~\cite{puhlhofer77}. The parameters needed for the
statistical description, i.e. the nuclear level densities and the barrier
transmission probabilities, are usually obtained from the study of evaporated
light particle spectra. In recent years, it has been observed that
statistical model calculations, using standard parameters, are unable to
predict satisfactorily the shape of the evaporated $\alpha$-particle energy 
spectra~\cite{viesti88,govil87,larana88,huizenga89,agnihotri93,govil98,bandyopadhyay99,govil00a,govil00b,mahboub01}
with the measured average energies of the $\alpha$-particles found to be much
lower than the corresponding theoretical predictions. In the present
calculations as well as in previous
studies~\cite{viesti88,govil87,huizenga89,agnihotri93,govil98,bandyopadhyay99,govil00a,govil00b,mahboub01},
the transmission coefficients of all competing evaporation channels, including
n, p, and $\alpha$ particle emission, are generated from published optical
model parameters for spherical nuclei. Several attempts have been made in the
past few years to explain LCP energy anomaly either by  changing the emission
barrier or by using a spin-dependent level density. The change in the emission
barriers and, correspondingly, the transmission probabilities affects the lower
energy part of the calculated evaporation spectra. On the other hand the
high-energy part of the spectra depends critically on the available phase space
obtained from the level densities at high spin. In hot rotating nuclei formed
in heavy-ion reactions, the level density at higher angular momentum is spin
dependent. The level density, $\rho(E,J)$, for a given angular momentum $J$ and
energy $E$ is given by the well known Fermi gas expression : 

\begin{equation}
\rho(E,J) = {\frac{(2J+1)}{12}}a^{1/2}
           ({\frac{ \hbar^2}{2 {\cal J}_{eff}}}) ^{3/2}
           {\frac{1}{(E-\Delta-T-E_J)^2} }exp(2[a(E-\Delta-T-E_J)]^{1/2})
\label{lev}
\end{equation}

where $a$ the level density parameter is constant and set equal to $a$ = A/8
MeV$^{-1}$ (A is the mass number), T is the ``nuclear" temperature, and
$\Delta$ is the pairing correction, E$_J$ = $\frac{ \hbar^2}{2 {\cal
J}_{eff}}$J(J+1) is the rotational energy, ${\cal J}_{eff}= {\cal J}_0 \times
(1+\delta_1J^2+\delta_2J^4)$ is the effective moment of inertia,  ${\cal J}_0$
= ${\frac{2}{5}}$AR$^{2}$ = ${\frac{2}{5}}$A$^{5/3}$r$_{0}^{2}$ is the rigid
body moment of inertia (r$_{0}$ is the radius parameter), and $\delta_1$ and
$\delta_2$ are the deformability parameters defined in
Refs.~\cite{viesti88,govil87,huizenga89}. 
 
The angular momentum distribution used in the statistical model calculations
depends on diffusivity parameter $\Delta$L and the critical angular momentum
for fusion L$_{cr}$. A fixed value of $\Delta$L = 1$\hbar$ is assumed for the
calculations. The L$_{cr}$ values were deduced based on observed complete
fusion cross section.  These values are shown in Table I for a number of
systems with 46 $\leq$ $A_{CN}$ $\leq$ 60. The same angular momenta value of
L$_{cr}$ = 34$\hbar$ is found for both the $^{28}$Si +
$^{28}$Si~\cite{dicenzo81,nagashima86,vineyard90} and $^{30}$Si +
$^{30}$Si~\cite{dumont85,bozek86} fusion reactions. Otherwise we have used the
{\sc CASCADE} parameters for the nickel isotopes (see Table 8 of
Ref.~\cite{bozek86}). The only parameters adjusted in the calculations were
those directly associated with the system deformation, $\delta_1$ and
$\delta_2$, the so-called deformability parameters. 

In the present analysis we have chosen to follow the procedure proposed by
Huizenga and collaborators~\cite{huizenga89}. No attempt was made to modify the
transmission coefficients since it has been shown that the effective barrier
heights are fairly insensitive to the nuclear deformation~\cite{huizenga89}. On
the other hand, by changing the deformability parameters $\delta_1$ and
$\delta_2$ one can simulate the spin-dependent level
density~\cite{viesti88,govil87,huizenga89}. The {\sc CACARIZO} calculations
have been performed using two sets of input parameters : one with a standard
set of the rotating liquid drop model~\cite{puhlhofer77} (RLDM) (parameter {\bf
set A}), consistent with the deformation of the finite-range rotating liquid
drop model~\cite{sierk86} (FRLDM), and another with a spin-dependent moment of
inertia and larger values for the deformability parameters (parameter {\bf set
B}). The RLDM parameter {\bf set A} with small values given to deformability
parameters ($\delta_1$ = 7.6 x 10$^{-6}$ and $\delta_2$ = 6.7 x 10$^{-8}$)
produces a yrast line very close to the FRLDM predictions, as shown for example
for the neighbouring $^{59}$Cu nucleus in Ref.~\cite{viesti88} (see Fig.~1 of
Ref.~\cite{viesti88}). The final values of the deformability parameters,
$\delta_1$ = 1.2 x 10$^{-4}$ and $\delta_2$ = 1.1 x 10$^{-7}$ given in Table I
for the parameter {\bf set B}, yield a significant lowering of the
corresponding FRLDM yrast line. They have been chosen in order to reproduce the
exclusive data rather than the inclusive data (although they have almost
identical spectral shapes especially in the high-energy region), the latter
being possibly influenced for low-energy LCp's by small non-statistical
components resulting from inelastic collisions or breakup processes that are
not accounted for in the statistical model calculations. 

The dashed lines in Fig.~3 show the predictions of {\sc CACARIZO} for $^{28}$Si
+ $^{28}$Si using the parameter {\bf set A} consistent with FRLDM
deformation~\cite{sierk86}. It is clear that the average energies of the
measured $\alpha$-particle spectra are lower than those predicted by these
statistical model calculations. The same observation can be made in Fig.~5 for
the inclusive $^{30}$Si + $^{30}$Si data \cite{larana88} which have been
analysed with the same parameters as used for the $^{28}$Si + $^{28}$Si
reaction (the corresponding inclusive $^{28}$Si + $^{28}$Si data are also
displayed in Fig.~6 for the sake of comparison).  The solid lines of Figs.~3, 5
and 6 show the predictions of {\sc CACARIZO} using the increased values of the
deformability parameters (see parameter {\bf set B} given in Table I). The
agreement is considerably improved. For instance in Figs.~3 the shapes of the
exclusive $\alpha$-particle energy spectra are very well reproduced for
$^{28}$Si + $^{28}$Si with parameter {\bf set B} (solid lines) including the
deformation effects. Furthermore and despite the fact that, for the inclusive
$\alpha$-particle spectra of the $^{28}$Si + $^{28}$Si reactions of Fig.~6,
there appears to be an excess of yield in the subbarrier energy data, the
chosen parametrization of the moment of inertia does also a fairly good job. It
may be mentionned that in the case of protons, as they carry away less angular
momentum than $\alpha$-particles, their calculated energy spectra do not shift
as the spin-dependent parametrization of the moment of inertia is introduced.
The statistical model results using the two parameter sets reproduce equally
well the experimental velocity spectra and angular correlations. The
statistical model calculations displayed for protons on Figs. 1, 2, and 4 have
been performed with parameter {\bf set B} (solid lines of Fig.4) including the
deformation effects (calculations with parameter {\bf set A} are not
displayed). The deformability parameters for the other systems, given in Table
I, were extracted using the same approach of fitting
procedures~\cite{viesti88,agnihotri93,mahboub01,nicolis89} that was employed in
the present work for the $^{28}$Si + $^{28}$Si and $^{30}$Si + $^{30}$Si
reactions. 

The {\sc CACARIZO} predictions shown in Figs.~1 and 2, also performed with the
parameter {\bf set B} given in Table I, reproduce the maxima of the inclusive
and exclusive invariant cross sections quite well. This confirms that most of
the yields have a statistical origin. This is consistent with the experimental
alpha-to-proton ratio R$_{\alpha/p}^{exp}$ = 0.40 $\pm$ 0.06 which value is
better predicted by calculations using parameter {\bf set B}
(R$_{\alpha/p}^{set-B}$ = 0.39) than calculations using parameter {\bf set A}
(R$_{\alpha/p}^{set-A}$ = 0.48). Parameter {\bf set A} overestimate the
alpha-to-proton ratio mainly because of first chance $\alpha$-particle.
Parameter {\bf set B} allows the emission of more nucleons in the cascade and
the average emission step for $\alpha$-particles occurs later in the cascade.
Thus the velocity plots, the spectral shapes, the angular correlations and
relative cross sections of $\alpha$-particle and proton emission are all
reproduced correctly. On the other hand, the discrepancies observed at the most
negative angles (between $\Theta^{LCP}_{lab}$ = -30$^\circ$ and -110$^\circ$)
of the in-plane angular correlations of Fig.~4, for protons and even more for
$\alpha$-particles, are sometimes difficult to be understood as already
stressed in Refs.~\cite{mahboub01,vineyard94}. The same disagreement is present
with the calculations using parameter {\bf set A}. However, it is seen that the
shapes of the experimental angular correlations are well reproduced by the
statistical theory for the positive angles, although a very small shift of
$\Delta$$\Theta^{LCP}_{lab}$ = 10-15$^{\circ}$ may improve the comparison. 

In addition, it is interesting to note that for the $^{30}$Si + $^{30}$Si
reaction the relative multiplicities of nucleons and $\alpha$-particles deduced
from the experimental data~\cite{bozek86} (M$_{\large n}^{exp}$ =
0.469$\pm$0.035, M$_{\large p}^{exp}$ = 0.343$\pm$0.035, and M$_{\large
\alpha}^{exp}$ = 0.188$\pm$0.035) are in better agreement with the calculations
using the parameter {\bf set B} (M$_{\large n}^{def}$ = 0.484, M$_{\large
p}^{def}$ = 0.312, and M$_{\large \alpha}^{def}$ = 0.204) than those using the
parameter {\bf set A} (M$_{\large n}^{\small LDM}$ = 0.459, M$_{\large
p}^{\small LDM}$ = 0.293, and M$_{\large \alpha}^{\small LDM}$ = 0.248). 

As a whole, the present statistical model calculations describe rather well all
the measured observables for both the $^{28}$Si + $^{28}$Si and $^{30}$Si +
$^{30}$Si reactions, in contrast to other recent studies which have needed
extra dynamical effects in the evaporative processes ~\cite{govil00a,govil00b}.

In order to better appreciate the magnitude of the possible deformation effects
which are suggested by our choice statistical model approach, one may express
the effective moment of inertia as 
${\cal J}_{eff}$ = ${\frac{2}{5}}$MR$^{2}$ = ${\frac{1}{5}}$M(b$^{2}$+a$^{2}$)
with the volume conservation condition: V = ${\frac{4}{3}}\pi$abc,
where b and a are the major and minor axis, and c is the rotational axis
of the compound nucleus.
In the case of an oblate shape a = b and ${\cal J}_{eff}$ =
${\frac{2}{5}}$Ma$^{2}$ 
and  V = ${\frac{4}{3}}\pi$a$^{2}$c. The axis ratio is equal to 
$\delta$ = a/c = (1+$\delta_1$J$^{2}$+$\delta_2$J$^{4}$)$^{3/2}$.
In the case of a prolate shape a = b and ${\cal J}_{eff}$ =
${\frac{1}{5}}$M(b$^{2}$+a$^{2}$) and  V = ${\frac{4}{3}}\pi$a$^{2}$b. We
obtain the equation : 1+(3-$\gamma$)x+3x$^{2}$+x$^{3}$ = 0 with x = $\left(
{\frac{b}{a}}\right)^{2}$ = ${\delta}^{2}$ and $\gamma$ = 8( 
1+$\delta_1$J$^{2}$+$\delta_2$J$^{4}$)$^{3}$.
The quadrupole deformation parameter $\beta$ is equal to $\beta$ =
${\frac{1}{\sqrt{5\pi}}}({\frac{4}{3}}\delta+{\frac{2}{3}}\delta^{2}+{\frac{2}{3}}\delta^{3}+{\frac{11}{18}}\delta^{4})$. 

The minor to major axis ratios b/a and the values of the quadrupole deformation
parameter $\beta$ are shown for the systems tabulated in Table 1. These 
quantities have been extracted from the fitted deformability parameters by
assuming either a symmetric oblate shape or a symmetric prolate shape,
respectively, with sharp surfaces~\cite{huizenga89}. All of the symmetric or
near symmetric systems~\cite{agnihotri93,mahboub01,nicolis89}, for which the
main parameters are displayed in the Table, appear to favor relatively large
deformations. This can be contrasted with what is found for the asymmetric
$^{12}$C + $^{45}$Sc reaction~\cite{govil00b}, where a standard statistical
model calculation~\cite{govil00b} is found to work well. The same conclusion
was reached for the $^{28}$Si + $^{51}$V reaction~\cite{govil98}, another
asymmetric system, and was confirmed for the very asymmetric system $^{16}$O +
$^{54}$Fe~\cite{govil00a}. 

From this systematic analysis, it can be observed that the magnitude of the
quadrupole deformation that can be deduced for both $^{28,30}$Si + $^{28,30}$Si
reactions are rather large : their parameter values are similar to the value
obtained for the $^{32}$S + $^{27}$Al reaction~\cite{viesti88,nicolis89} but
larger than the value corresponding to the $^{28}$Si + $^{27}$Al
reaction~\cite{agnihotri93}. This leads to the striking conclusion that highly
stretched configurations are required to account for the observed
$\alpha$-particle evaporation spectra. For $^{28}$Si + $^{28}$Si the value of
$\beta$ $\approx$ 0.5 found for the quadrupole deformation parameter is
consistent with the recent observation of very deformed bands in the
doubly-magic $^{56}$Ni nucleus by standard $\gamma$-ray spectroscopy
methods~\cite{rudolph99}. 

\newpage

\section{CONCLUSION}

To summarize, the properties of the light charged particles emitted in the
$^{28}$Si + $^{28}$Si reaction at the bombarding energy E$_{lab}$($^{28}$Si) =
112 MeV, which corresponds to the $^{56}$Ni excitation energy of the
conjectured J$^{\pi}$ = 38$^{+}$ quasimolecular
resonance~\cite{nouicer99,beck01}, have been investigated using the statistical
model. The measured observables such as velocity distributions, energy spectra,
in-plane angular correlations, and multiplicities are all reasonably well
described by the Monte Carlo {\sc CASCADE} calculations requiring
spin-dependent level densities. The magnitude of the adjustments in the yrast
line position suggests deformation effects at high spin for the $^{56}$Ni
composite system in agreement with very recent $\gamma$-ray spectroscopy data
obtained at much lower spins~\cite{rudolph99}. The extent to which the resonant
behaviour is responsible to the observed deformation is still an open question.
To resolve this issue, more exclusive data are needed at off-resonance
energies. The extracted deformability parameters for $^{28}$Si + $^{28}$Si are
consistent with the re-analysis of previously published $^{30}$Si + $^{30}$Si
data. We conclude that mass-symmetric systems do favor the onset of strong
deformation effects at high angular momenta for highly stretched configurations
without the need of recently speculated dynamical
effects~\cite{govil00a,govil00b} to explain the $\alpha$-particles energy
spectra. Work is in progress, including out-of-plane correlations measurements,
to study the same reactions with the ICARE multidetector array at even higher
bombarding energies~\cite{rousseau01b}. 

\newpage

\centerline{\bf ACKNOWLEDGMENTS }

\vskip 0.8cm

This work is based upon the Ph.D.~Thesis of M.~Rousseau, Universit\'e Louis
Pasteur, Strasbourg, 2000. We would like to thank the staff of the VIVITRON for
providing us with good stable beams, J.~Devin and C.~Fuchs for the excellence
support in carrying out these experiments. Particular appreciation to M.A.
Saettel for preparing the targets, and to J.P.~Stockert and A.~Pape for
assistance during their RBS measurements. One of us (M.R.) would like to
acknowledge the Conseil R\'egional d'Alsace for the financial support of his
Ph.D.~Thesis work. This work was sponsored by the French CNRS/IN2P3, and by
CNRS/NSF and CNRS/CNpq collaboration programs, and also in part by the U.S. DOE
under Grant No. DE-FG03-96-ER40981.

\hspace{-2cm}
\renewcommand{\baselinestretch}{1.2}
\begin{table}[h!]
\hspace{-2cm}
\begin{tabular}{|c|c|c|c|c|c|c|c|c|}
\hline
Reaction & C.N. & Energy (MeV)& $L_{cr}$ ($\hbar$) & $\delta_1$ & $\delta_2$ &
b/a & $\beta$ & References\\
\hline
$^{28}$Si+$^{27}$Al & $^{55}$Co  &  150 & 42 & 1.8$\cdot$10$^{-4}$ &
1.8$\cdot$10$^{-7}$ & 1.2/1.3 & -0.44/0.46 & \cite{agnihotri93} \\
\hline
$^{28}$Si+$^{28}$Si & $^{56}$Ni & 112 & 34 &
1.2$\cdot$10$^{-4}$ & 1.1$\cdot$10$^{-7}$ & 1.5/1.6 & -0.48/0.59 & This
work \\
\hline
$^{35}$Cl+$^{24}$Mg & $^{59}$Cu  & 260 & 37 & 1.1$\cdot$10$^{-4}$ &
1.3$\cdot$10$^{-7}$ & 1.6/1.7 & -0.50/0.51 & \cite{mahboub01}\\
\hline
$^{32}$S+$^{27}$Al & $^{59}$Cu  & 100 - 150 & 27 - 39 & 2.3$\cdot$10$^{-4}$ &
1.6$\cdot$10$^{-7}$ & 1.4 - 2.0 & -0.46/0.53 & \cite{viesti88} \\
\hline 
$^{32}$S+$^{27}$Al & $^{59}$Cu  &  100 - 150 & 27 - 42 & 1.3$\cdot$10$^{-4}$
& 1.2$\cdot$10$^{-7}$ & 1.5 - 2.2 & -0.48/0.54 & \cite{nicolis89} \\
\hline
$^{30}$Si+$^{30}$Si & $^{60}$Ni & 120 & 34 &
1.2$\cdot$10$^{-4}$ & 1.1$\cdot$10$^{-7}$ & 1.6/1.7 & -0.49/0.50 & This
work \\
\hline
\end{tabular}

\vskip 0.7cm

\renewcommand{\baselinestretch}{1.}
\caption{\label{TABLE I}\textit{\sl Typical quantities of the evaporation
calculations performed using the statistical model code {\sc CACARIZO}. The
deformability parameters are taken either from the parameter {\bf set B} for
the systems studied in the present work or from similar fitting procedures for
the other systems studied in the literature. The minor to major axis ratios 
b/a and the quadrupole deformation $\beta$ values (for a symmetric oblate shape
and a symmetric prolate shape, respectively) have been deduced from equations
discussed in the text. Note that the $\beta$ values given for
$^{32}$S + $^{27}$Al have been extracted assuming the $L_{cr}$ as extracted at
the highest bombarding energy.}} 
\end{table}

\newpage

\begin{figure}
\centering
\epsfig{figure=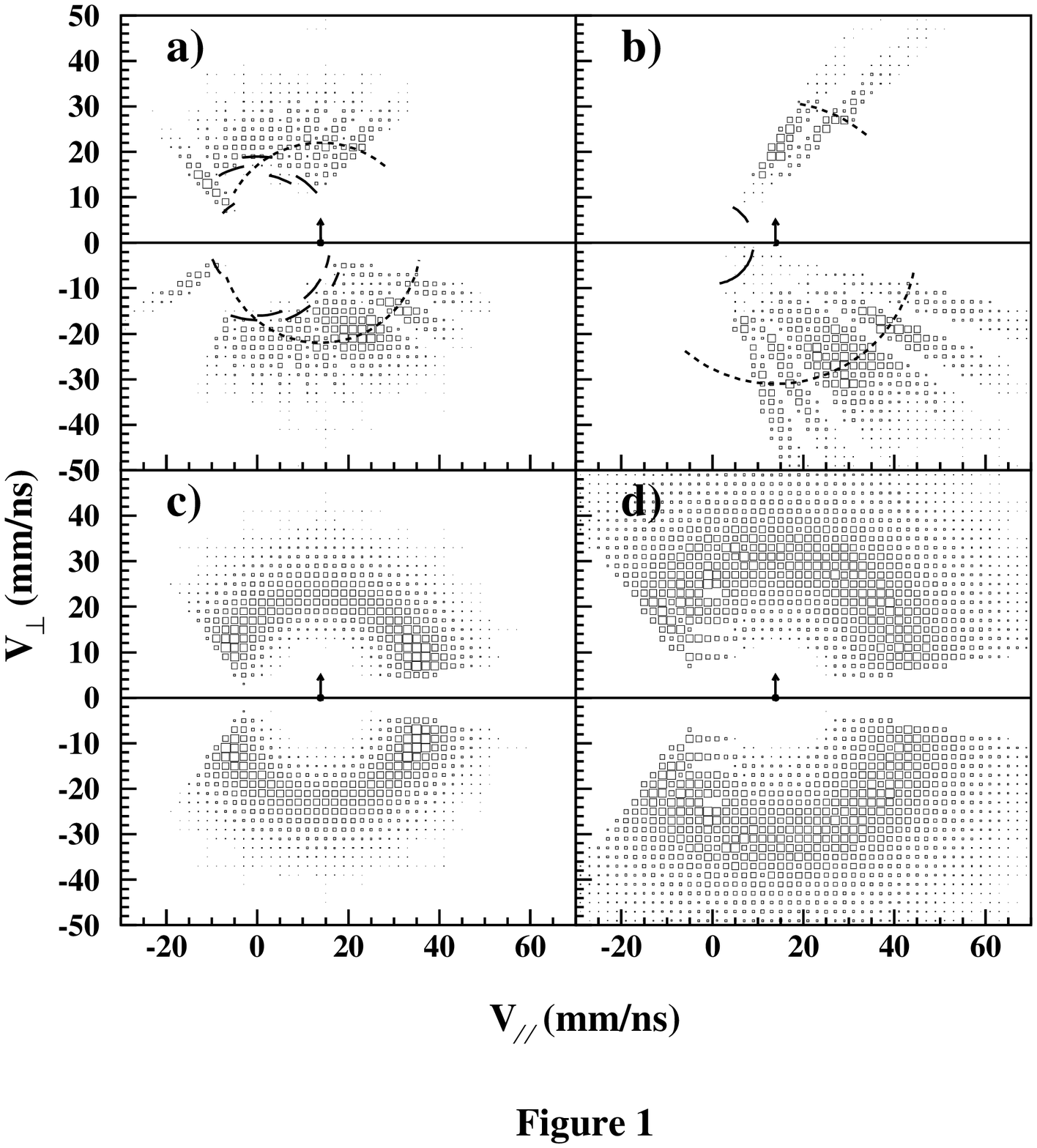,width=15.0cm}

\vspace{1cm}

\caption{ Two-dimensional scatter plots of Galilean invariant cross-sections
(d$^{2}\sigma$/d$\Omega$dE)p$^{-1}$c$^{-1}$ of $\alpha$-particles (a) and
protons (b), respectively, in the (V$_{\parallel}$,V$_{\perp}$) plane for the
$^{28}$Si(112 MeV) + $^{28}$Si reaction. The experimental detector thresholds
are drawn along the laboratory angles of each telescope. (c) and (d) are the
corresponding statistical model calculations discussed in the text. The dashed
circular arc are centered on the velocity of the center of mass indicated by
the arrows.} 
\end{figure}

\begin{figure}
\centering
\epsfig{figure=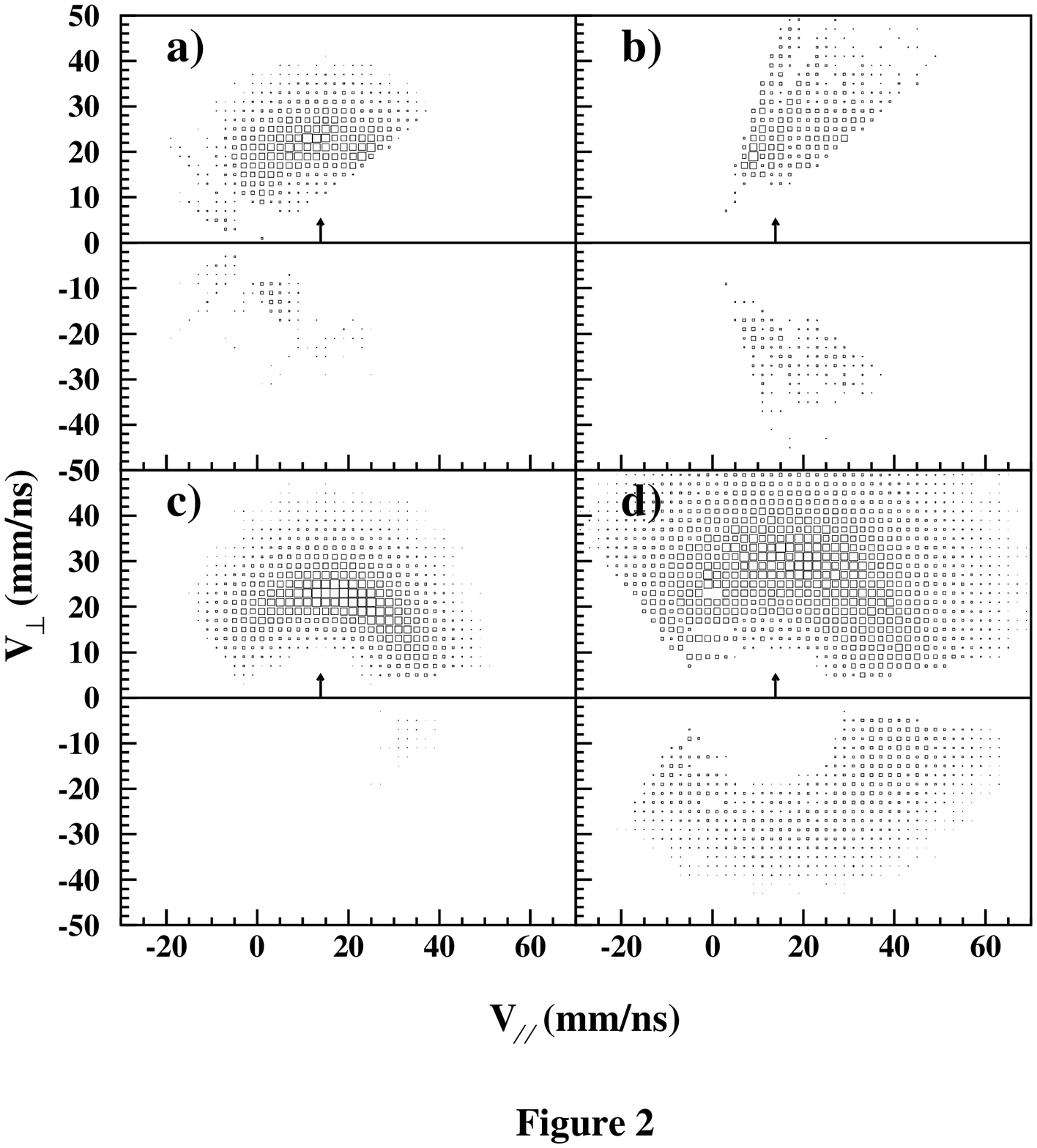,width=15.0cm}

\vspace{1cm}

\caption{Two-dimensional scatter plots of exclusive Galilean invariant
cross-sections (d$^{2}\sigma$/d$\Omega$dE)p$^{-1}$c$^{-1}$ of
$\alpha$-particles (a) and protons (b), respectively, measured in coincidence
with all ER's (20 $\leq$ Z $\leq$ 25) identified in a IC detector located at
$\Theta^{ER}_{lab}$ = -15$^{\circ}$, as plotted in the
(V$_{\parallel}$,V$_{\perp}$) plane for the $^{28}$Si(112 MeV) + $^{28}$Si
reaction. (c) and (d) are the corresponding results of the statistical model
calculations discussed in the text. The arrows indicate the center-of-mass
velocity.} 
\end{figure}

\begin{figure}
\centering
\epsfig{figure=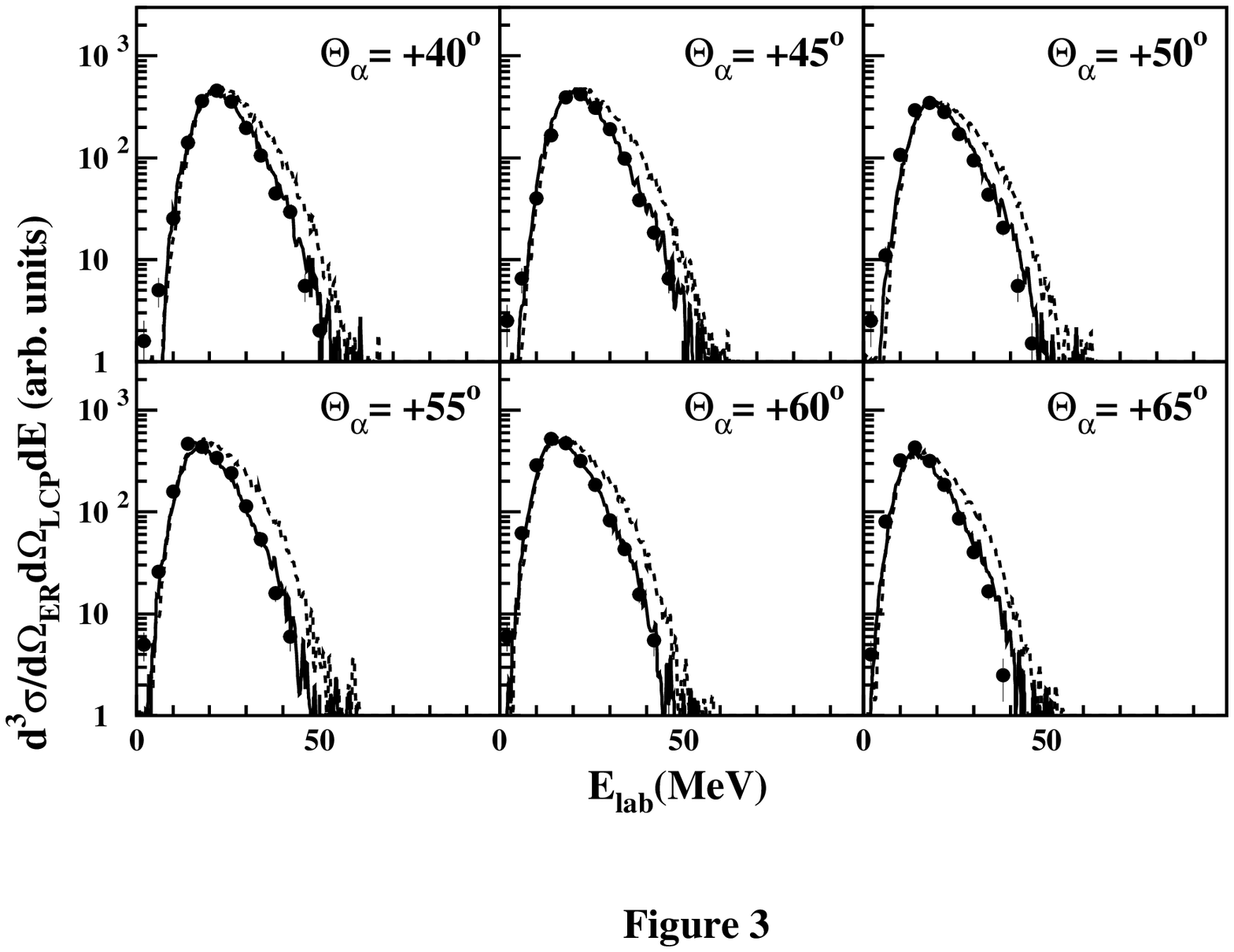,width=15.0cm}

\vspace{1cm}

\caption{ Exclusive energy spectra of $\alpha$-particles in coincidence with
all ER's (20 $\leq$ Z $\leq$ 25), identified in a IC detector located at
$\Theta^{ER}_{lab}$ = -15$^{\circ}$, produced in the $^{28}$Si(112 MeV) +
$^{28}$Si reaction. The experimental data are given in relative units by the
solid points with error bars visible when greater than the size of the points.
The dashed and solid lines are the results of statistical model calculations
using parameter {\bf set A} and {\bf set B}, respectively, as discussed in the
text.}
\end{figure}

\begin{figure}
\centering
\epsfig{figure=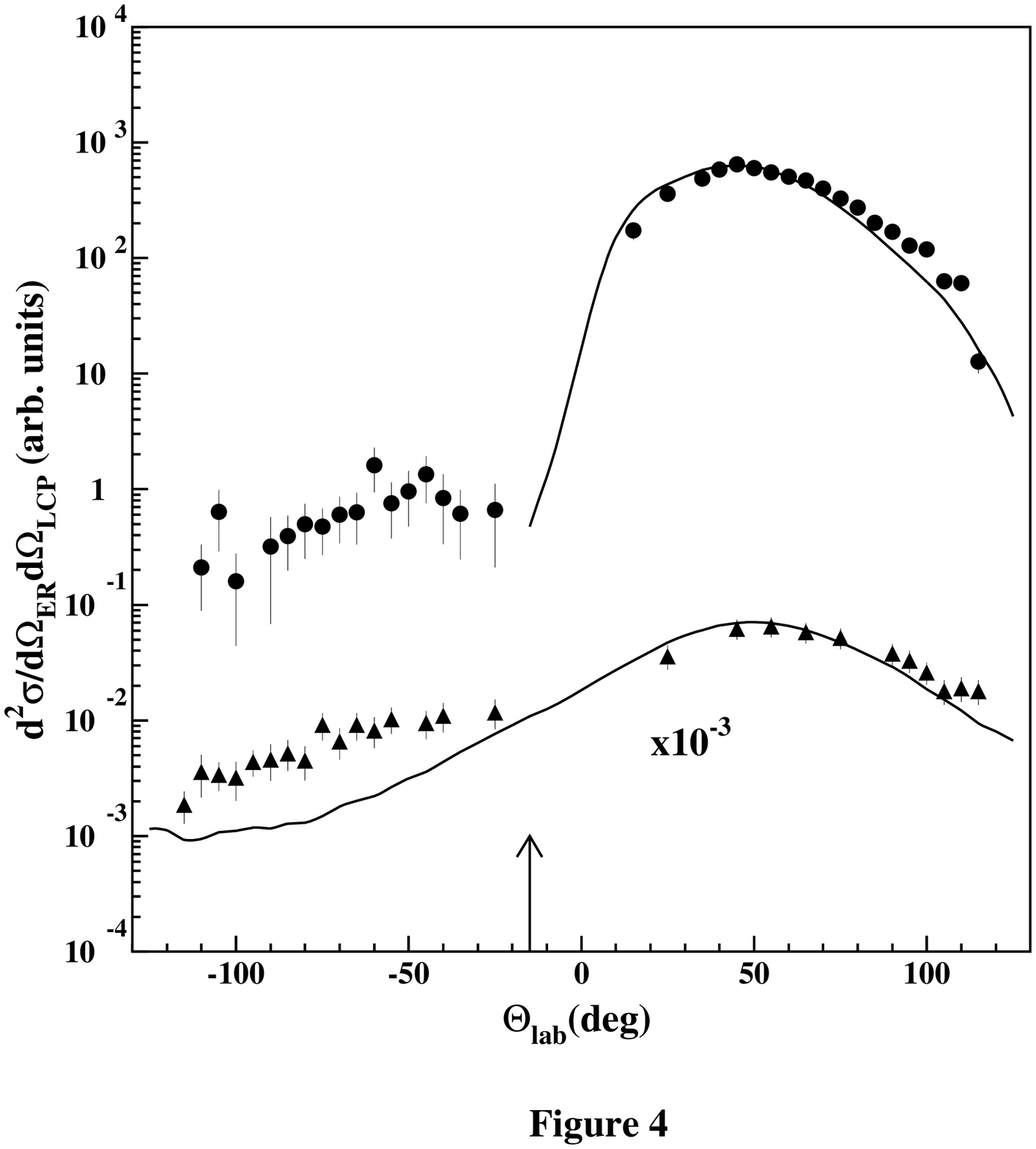,width=15.0cm}

\vspace{1cm}

\caption{ In-plane angular correlation (-115$^{\circ}$ $\leq$
$\Theta^{LCP}_{lab}$ $\leq$ +115$^{\circ}$) of $\alpha$-particles (circles) and
protons (triangles) in coincidence with all ER's (20 $\leq$ Z $\leq$ 25)
produced in the $^{28}$Si(112 MeV) + $^{28}$Si reaction and detected at
$\Theta^{ER}_{lab}$ = -15$^{\circ}$ as shown by the arrow. The solid curves are
the results of statistical model calculations using parameter {\bf set B} as
discussed in the text. }
\end{figure} 

\begin{figure}
\centering
\epsfig{figure=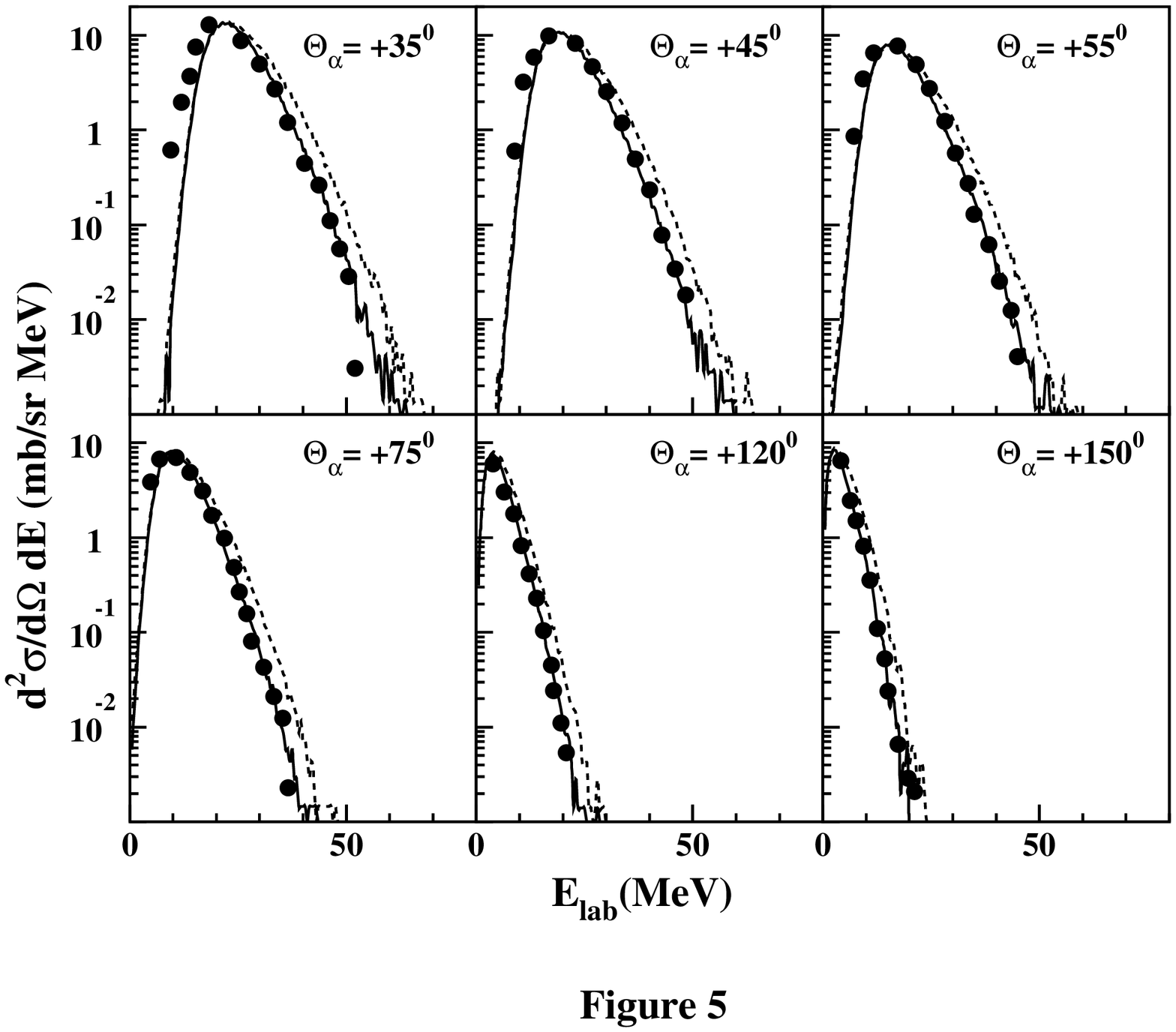,width=15.0cm}

\vspace{1cm}

\caption{Energy spectra of $\alpha$-particles produced in the $^{30}$Si(120
MeV) + $^{30}$Si reaction. The data taken from ref.\protect\cite{larana88} (solid
points) are compared to the results of the statistical model calculations
(solid and dashed lines correspond to parameter {\bf set B} and {\bf set A},
respectively) discussed in the text.} 
\end{figure}

\begin{figure}
\centering
\epsfig{figure=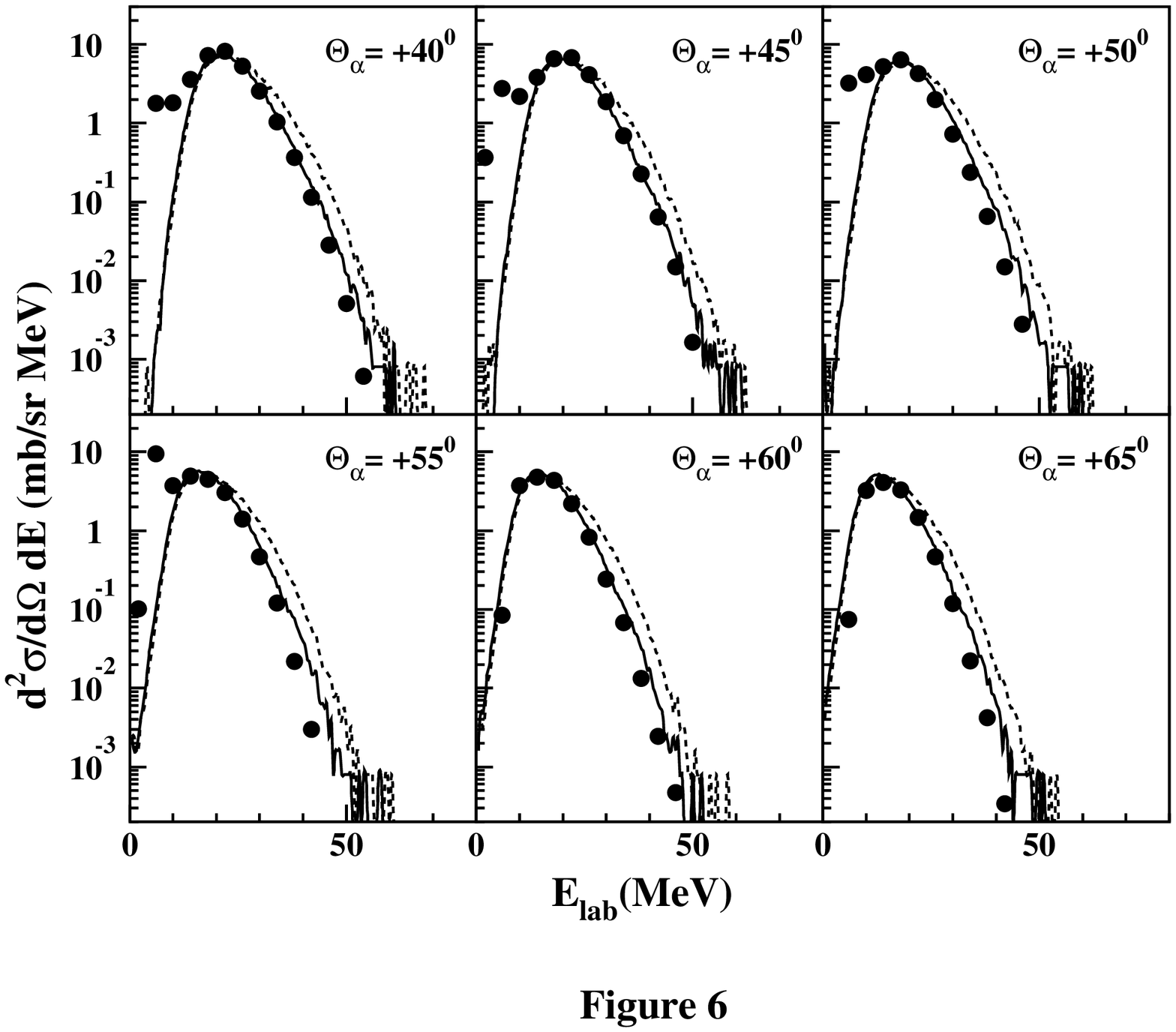,width=15.0cm}

\vspace{1cm}

\caption{ Energy spectra of $\alpha$-particles produced in the $^{28}$Si(112
MeV) + $^{28}$Si reaction. The inclusive cross sections are given as absolute
values by the solid points with error bars visible when greater than the size
of the points. The dashed and solid lines are the results of statistical model
calculations using parameter {\bf set A} and {\bf set B}, respectively, as
discussed in the text. }
\end{figure}

\end{document}